# Stuffed Rare Earth Garnets

Chen Yang[1,2], Lun Jin[1], Weiwei Xie[3] and Robert J. Cava[1*]

[1] Department of Chemistry, Princeton University, Princeton, New Jersey, 08544

[2] Department of Electrical and Computer Engineering, Princeton University, Princeton, New Jersey, 08544

[3] Department of Chemistry, Michigan State University, East Lansing, Michigan, 48824

*corresponding author's email: rcava@princeton.edu

**Abstract**

We report the synthesis and magnetic characterization of stuffed rare earth gallium garnets, $RE_{3+x}Ga_{5-x}O_{12}$ (RE=Lu, Yb, Er, Dy, Gd), for $x$ up to 0.5. The excess rare earth ions partly fill the octahedral sites normally fully occupied by $Ga^{3+}$, forming disordered pairs of corner-shared face-sharing magnetic tetrahedra. The Curie-Weiss constants and observed effective moments per rare earth are smaller than are seen for the unstuffed gallium garnets. No significant change in the field-dependent magnetization is observed but missing entropy is seen when integrating the low-temperature heat capacity to 0.5 K.

**Introduction**

The rare earth oxide garnets ($A_3B_5O_{12}$) and pyrochlores ($A_2B_2O_7$) are two families of materials that have been employed to study magnetism ([1–4]). The hyperkagome net in garnets, a corner-sharing triangular net **(Figure 1a)** ([5]), when formed by rare earth ions, leads to magnetic properties that are said to be spin-liquid-like ([2,3,6,7]), and the magnetic tetrahedra in pyrochlores can lead to spin-ice-like magnetic properties ([4]). "Stuffed" rare earth pyrochlores have been synthesized and studied previously ([1,8,9]), as have garnets where different rare earths have been mixed together ([10–15]). Unlike the pyrochlores, the non-magnetic sites in garnets are not only octahedrally coordinated to oxygen (2 per formula unit) but are also tetrahedrally coordinated to oxygen (3 per formula unit), such that the formula is better written structurally as $A_3B_2B'_3O_{12}$. In pyrochlore-based $Ho_2Ti_{2-x}Ho_xO_{7-x/2}$, for example, where $0 \leq x \leq 0.67$, the stuffed $Ho^{3+}$ goes on the $Ti^{4+}$ octahedral site. The stuffed pyrochlores gradually undergo a structural transition from purely corner-sharing magnetic tetrahedra to the fluorite structure, which results in a disordered array of both edge-sharing and corner-sharing magnetic tetrahedra. A stuffing concentration of up to $x = 0.67$ succeeds for stuffed pyrochlores of the type $Ln_2Ti_{2-x}Ln_xO_{7-x/2}$ for small rare earths, but though also successful, the stuffing is more difficult for larger rare earth ions. The structural disorder does not affect the temperature-dependent magnetic susceptibility per rare earth but modifies the field-dependent magnetization and magnetic entropy ([8,9,16,17]).

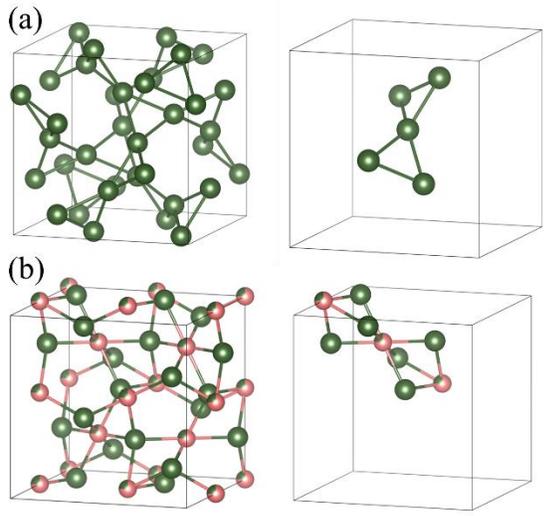

**Figure 1. Crystal structure of the gallium garnets.** (a) The hyperkagome lattice formed by Er in stoichiometric $Er_3Ga_5O_{12}$. On the right, one pair of corner-sharing triangles is shown for clarity. (b) The face-sharing and corner-sharing tetrahedra of magnetic ions formed when Er occupies the dodecahedral and octahedral sites in stuffed gallium garnets. On the right, one pair of tetrahedra is shown for clarity. View the green Er atoms form a shared face, and two vertices are the pink/green atoms. The middle pink/green atom is the shared corner of two face-sharing tetrahedra.

Here we report the synthesis of a different family of materials, the stuffed gallium garnets, nominally written as $RE_{3+x}Ga_{5-x}O_{12}$, for Gd and smaller rare earths. Under the conditions employed, Gd is the largest, rare earth ion that can be stuffed into its gallium garnet, and for the Fe- and Al-based garnets no stuffing was possible under our conditions. Optical studies of rare-earth-rich aluminum garnet have been reported, however, ([18]). Following our previous work on Er-excess gallium garnet ([19]), the present work studies whether stuffing other rare earth gallium garnets is possible and how the structure and magnetic properties of the resulting material are affected.

Unlike the case for pyrochlores, the highest stuffing concentration that we found for the garnets depends on the rare earth radius. The doped rare earth ions enter the 16a octahedral site of space group *Ia-3d*, which is normally fully occupied by $Ga^{3+}$, causing magnetic disorder **(Figure 1b)**: the stuffed garnets made here are isostructural with their stoichiometric variants. Very little change was found in the magnetic moment and the magnetization per rare earth ion. Significant changes in the low temperature heat capacity reveal, however, that the stuffing significantly influences magnetic frustration. We conclude that stuffed gallium garnets can be good comparisons to the stuffed pyrochlores for studying rare earth magnetism.

**Experimental**

The $RE_{3+x}Ga_{5-x}O_{12}$ (RE=Lu, Yb, Er, Dy, Gd) garnets were synthesized using a conventional solid-state method. The rare earth oxides (Thermo scientific 99.99%) were dried overnight at 900 °C and the gallium oxide (Sigma Aldrich 99.99%) was dried at 400 °C. The starting reagents were measured in molar ratios and mixed well in a mortar. The reaction was carried out in air for 2 weeks at 1500 °C (Sentro tech ST-1600C-445) with intermediate grindings. Powder X-ray diffraction (PXRD) collected on Brucker D8 FOCUS diffractometer with Cu Kα radiation ($\lambda_{K\alpha}$ = 1.5406 Å) was used to determine the phase purity. The TOPAS program suite was used to perform LeBail fits on all the PXRD patterns. Single crystal Xray diffraction was carried out using a XtalLAB Synergy, Dualflex, Hypix single crystal diffractometer with Mo K alpha radiation. A detailed description of the single crystal refinement procedure was presented in our previous paper ([20]).

Magnetic and heat capacity measurements were carried out on a Quantum Design PPMS. DC magnetic susceptibility (χ) was performed at temperatures between 300 K and 1.8K using a vibrating sample magnetometer (VSM). The applied field, H, was kept at 0.1T for all zero-field cooling (ZFC) measurements. The low temperature heat capacity measurements were performed in the same instrument between 0.4 K and 10 K.

**Results and discussion**

The stuffed samples remain isostructural to the stoichiometric garnets; **Figures 2a** and **2b** present the PXRD patterns of the Yb and Gd series as examples. Doping of excess Yb was successful until *x*=0.5 while for *x*= 0.6, the $Yb_2O_3$ impurity present indicates that the stuffing limit was reached. Successful stuffing from *x*=0.1 to *x*=0.5 was achieved for the Lu, Yb, and Er series.

However, the maximum amount of rare earth excess possible decreased for larger rare earths such as Dy and Gd. In the inset of **Figure 2a,** no peak shift was observed from $x=0.5$ to $x=0.6$ for the Yb series, while we could still see a minor peak shift from $x=0.2$ to $x=0.3$ for the Gd series. Though this could be evidence that more $Gd^{3+}$ can potentially be stuffed into the garnet, we could not make the Gd series free of $Gd_2O_3$. A temperature above 1500 °C melted the sample and longer heating time at 1500°C caused no change in the PXRD patterns. Therefore, we can safely claim the saturation limit for the Gd series is $x=0.2$ under the conditions employed here.

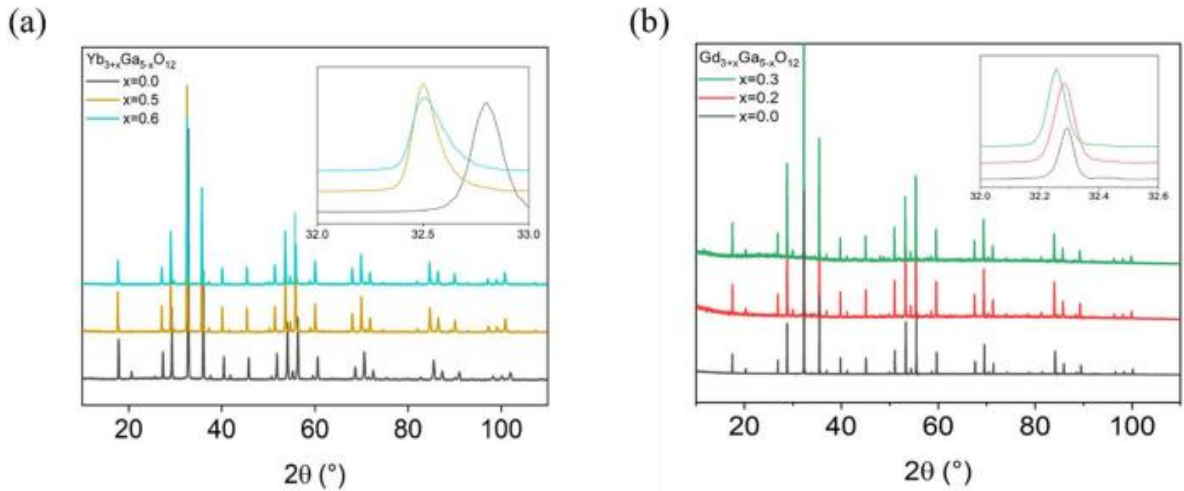

**Figure 2 . PXRD data and characterization of $RE_{3+x}Ga_{5-x}O_{12}$.** (a) The Yb PXRD patterns show stoichiometric (black), the most doped (yellow). Inset is a zoom-in of the 2θ range from 50°~60° showing an obvious peak shift from $x=0$ to $x=0.5$. The saturation is before $x=0.6$ because no peak shift was observed from $x=0.5$ to $x=0.6$. (b) The Gd PXRD patterns are stoichiometric (black), $x=0.2$ (red), and $x=0.3$ (green). The inset is a zoom-in of a selected peak in both panels.

The lattice parameters for all the samples are plotted in **Figure 3**. In **Figure 3a**, we plot the normalized lattice parameters against doping concentration. They all present a linear relationship between the doping concentration and lattice parameters. These data show that the maximum doping concentration was affected by the rare earth size; fewer rare earth ions could successfully substitute for the octahedral Ga with increasing rare earth ion size. In **Figure 3b**, we plot the lattice parameters for $x=0.0$ and $x=0.2$ versus the six-fold-coordination Shannon ionic radius for all the lanthanides studied. The linear increase observed is similar to what is observed for the stuffed pyrochlores ([9]).

Finally, in **Figure 3c**, we plot the normalized cell parameter versus the rare earth ionic radius to determine whether the lattice expansion was caused by rare earth concentration or by rare earth radii. We conclude that the lattice parameter is stretched by the stuffed rare earth ions, but not simply because a larger rare earth makes the lattice size larger. Further, more detailed structural comparisons of the pyrochlore and garnet systems may therefore be of interest.

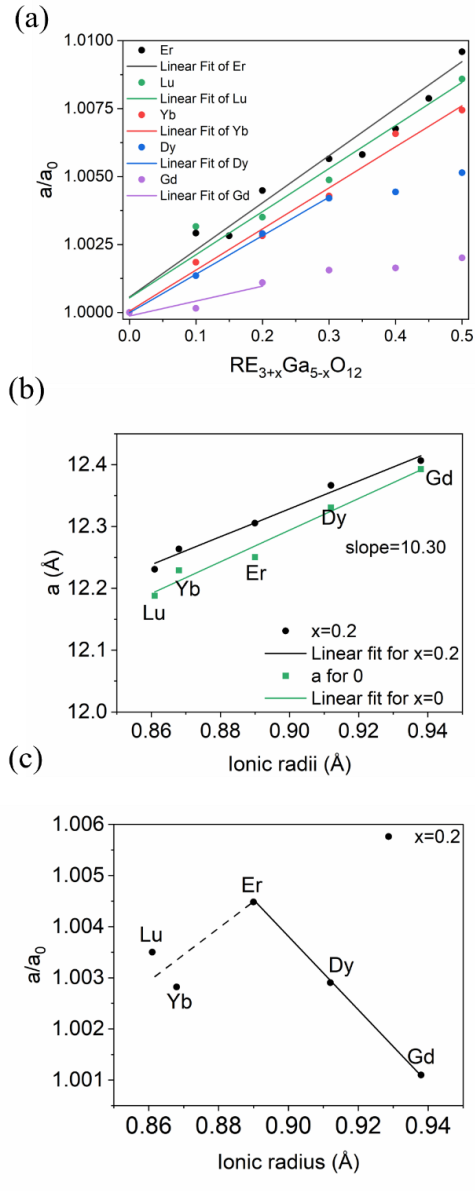

**Figure 3. Processed PXRD data.** (a) The normalized lattice parameters for all the samples synthesized, where $x = 0$ is taken as 1.0 (b) The lattice parameter for $RE_{3.2}Ga_{4.8}O_{12}$ when $x=0.0$ and $x=0.2$ versus the (octahedral) Shannon ionic radius for the lanthanide. (c) The normalized lattice parameter for $x = 0.2$ versus octahedral rare earth radius. Lines are guides to the eye.

We collected temperature-dependent magnetic susceptibility ($\chi$) and isothermal magnetization data for all the most highly stuffed $RE_{3+x}Ga_{5-x}O_{12}$ materials made (**Figures 4 and 5**). We fitted the Curie-Weiss (C-W) law to the data for the 2~5K to minimize crystal field effects (**Figure 4 and Figure S2**). A high-temperature 150~250K fit was also done. The C-W parameters are presented in **Table 1** and **Table S1**. No long-range ordering was observed in any of the samples within our temperature limits. The fitted Curie temperatures were negative, an indication of dominant anti-ferromagnetic interaction among the rare earth ions in all cases. The fitted effective moments for rare earth ions in the low temperature range are smaller than the free ion values.

Field-dependent zero-field cooling magnetic susceptibility was taken on $Er_{3.5}Ga_{4.5}O_{12}$ (**Figure S1**). No long-range ordering was observed across all magnetic fields employed. The temperature dependent behavior clearly indicates the presence of frustrated magnetism. This trend is consistent with the behavior of $Er_3Ga_5O_{12}$ previously reported ($^{21}$). The isothermal magnetization data in **Figure 5** were collected at 2K (black) and 300K (blue). The Yb-Gd samples behaved paramagnetically, and the Lu sample behaved diamagnetically ($Lu^{3+}$ has a filled 4$f$ shell) at room temperature, with a small paramagnetic tail due to impurity spins at low temperatures, as expected. Only Yb reached any clear magnetization saturation, while the rest of the magnetic rare earth ions may not do so within our measurable temperature and applied field range, although the data for the Gd sample may reveal some applied-field-induced transitions at 2 K. There is only a subtle difference in behavior for the stuffed and stoichiometric samples. For Er for example, the magnetic field inflection points of the M vs H curve for $Er_3Ga_5O_{12}$ and $Er_{3.5}Ga_{4.5}O_{12}$ are approximately 1.1T and 1.25T, respectively: our data suggest that a higher field is required to force a spin flip for the stuffed samples.

**Table 1. Curie-Weiss constants and magnetic moments for the stuffed end members of RE$_{3+x}$Ga$_{5-x}$O$_{12}$**

|  | Fit Range (K) | Fitted θ$_{cw}$ (K) | Experimental μ$_{eff}$ (μ$_B$/f.u.) | Expected μ$_{eff}$ (μ$_B$/f.u.) [22] | Δ of μ$_{eff}$ |
|---|---|---|---|---|---|
| Yb$_{3.5}$Ga$_{4.5}$O$_{12}$ | 2-5 | -0.55(5) | 3.292(1) | 4.53 | -27.3% |
|  | 150-250 | -91.09(3) | 5.068(7) |  | +11.9% |
| Er$_{3.5}$Ga$_{4.5}$O$_{12}$ | 2-5 | -1.08(1) | 7.062(7) | 9.58 | -26.3% |
|  | 150-250 | -12.91(6) | 10.057(4) |  | +13.3% |
| Dy$_{3.3}$Ga$_{4.7}$O$_{12}$ | 2-5 | -0.45(8) | 9.381(3) | 10.65 | -11.9% |
|  | 150-250 | -23.90(8) | 11.568(2) |  | +8.6% |
| Gd$_{3.2}$Ga$_{4.8}$O$_{12}$ | 2-5 | -0.56(1) | 6.098(1) | 7.94 | -23.2% |
|  | 150-250 | -5.15(3) | 8.421(5) |  | +6.1% |

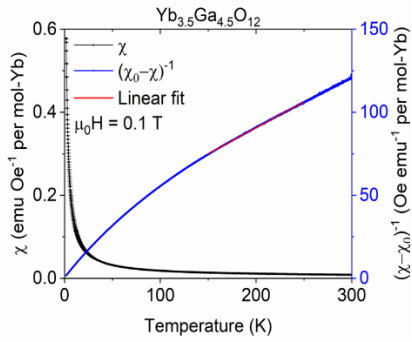
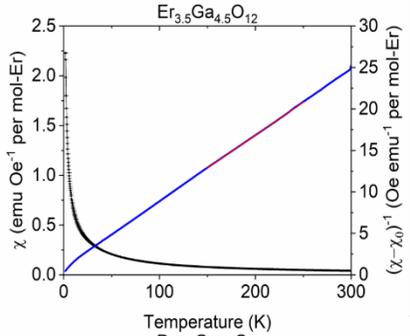
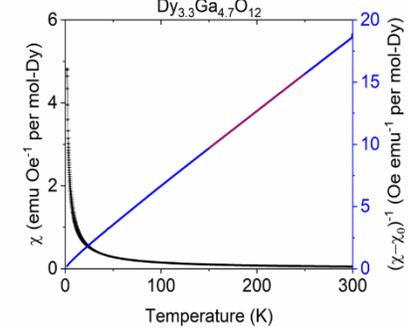
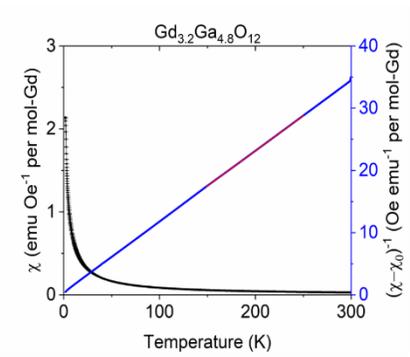
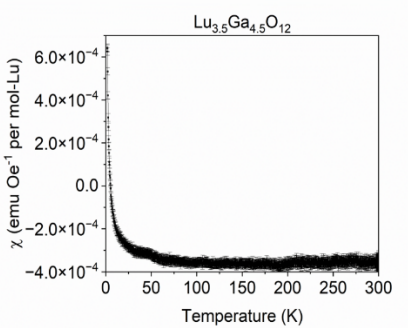

**Figure 4. Temperature-dependent magnetic susceptibility of selected stuffed garnets.** The red lines indicate the high temperature fits to the Curie-Weiss law.

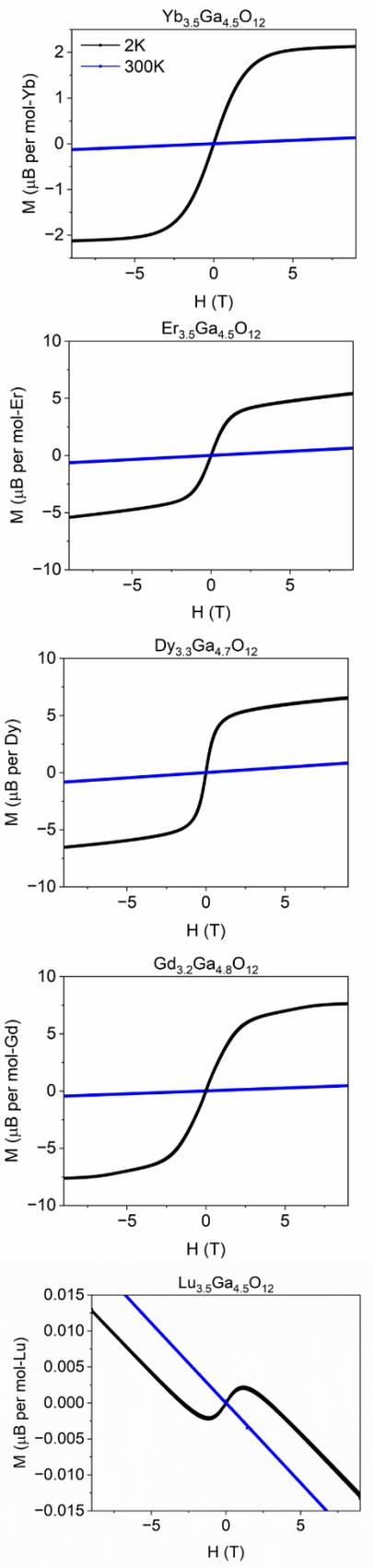

**Figure 5. Isothermal magnetization at 2K and 300K for representative stuffed garnets**.

The specific heat capacity data for $RE_{3.2}Ga_{4.8}O_{12}$ is useful for studying thermal and magnetic properties at temperatures between 0.5 and 1.8K. In the specific heat data under zero applied field, only the Er sample exhibits a long-range ordering transition, around 0.8K, in this temperature range while the rest of the magnetic rare earths presumably only display long-range order below 0.5K **(Figure 6)** ([2,23,24]) and **(Table 2).** Interestingly, $Gd_3Ga_5O_{12}$ (GGG) was reported to have an antiferromagnetic short-range transition temperature at $T_g$~0.8K under zero-field, which, judging from the broad peak, can also be a spin-glass-like transition ([2,23–25]). In our $Gd_{3.2}Ga_{4.8}O_{12}$ sample, a complete peak was not observed above 0.5K. The right-half-transition profile observed suggests that it has a narrower and shifted peak. This change in peak profile and position suggests a suppression of a spin-glass transition. A similar drastic change in peak profile and position was also observed when applying a 1T magnetic field on GGG and in the zero-field specific heat for the mixed rare earth garnet $Dy_{3-x}Gd_xGa_5O_{12}$ (DGGO) ([28]). In contrast, as shown in our previous report on Er series, the transition temperature of ~0.8K did not change for different $x$ values under zero applied field, and there was no clear change in the peak profile. A similar observation could not be drawn for Yb and Dy in our study because their transition temperature in zero-field specific heat is presumably below 0.5K. In Table 2, which compares stoichiometric and rare earth excess materials, the trend in ΔS is not clear. To support this table, we include **Figure S3**, which shows the magnetic entropy changes for the full Yb series. With the exception of Yb, the $x$=0.2 samples have a smaller saturated entropy per rare earth than the stoichiometric ones. In **Figure 7**, the specific heat data is shown, as collected under 0, 1T and 3T applied fields. The rare earth ions involved in this plot are Kramer's ions. (Kramer's ions can give doubly degenerate energy levels in a crystal field ([29])). In such systems, a Schottky-type anomaly is expected to broad and move to higher temperatures. In all the rare earth series studied, the peak broadened as expected. However, how the entropies for each rare earth react to applied magnetic fields are different. In some cases, the saturated entropies are close under 0T and 1T applied fields, while for Gd and Yb the saturated entropy under a 3T applied is around half of the amount that it is under 0T. How much the peak broadens under fields is also different.

For the Gd case, the integrated entropy saturates to a value in clear excess of Rln 2 suggesting that I is not an Ising ion in this system. Finally, we integrated the zero-field specific heat capacity data for the Er series both per total Er and per formula unit **(Figure S4)**. After comparing the data, we conclude that neither the absence of coupling between the two types of Er nor their strong coupling is a good description of these materials, though entropy is clearly missing no matter which scenario is considered; this leaves as an open question the nature of the magnetic rare earth coupling between the stoichiometric and stuffed sites in these materials. Hopefully, these unusual data can point out suitable directions for future expert study of the magnetism of stuffed garnets.

**Table 2. Comparison of magnetic entropy for *x*=0 and for the stuffed garnet at *x*=0.2.**

|  | Observed ΔS (J mol$^{-1}$ per mol-RE) | Stoichiometric observed ΔS (J mol$^{-1}$ per mol-RE) | % of excess ΔS For an Ising spin system. |
|---|---|---|---|
| Yb$_{3.2}$Ga$_{4.8}$O$_{12}$ | 2.6 | ~2.25 | 15.6% |
| Er$_{3.2}$Ga$_{4.8}$O$_{12}$ | 4.9 | ~6.0 ([20]) | -18.3% |
| Dy$_{3.2}$Ga$_{4.8}$O$_{12}$ | 2.6 | ~5 ([30]) | -18% |
| Gd$_{3.2}$Ga$_{4.8}$O$_{12}$ | 10 | 15.9 ([26]) | -37.1% |

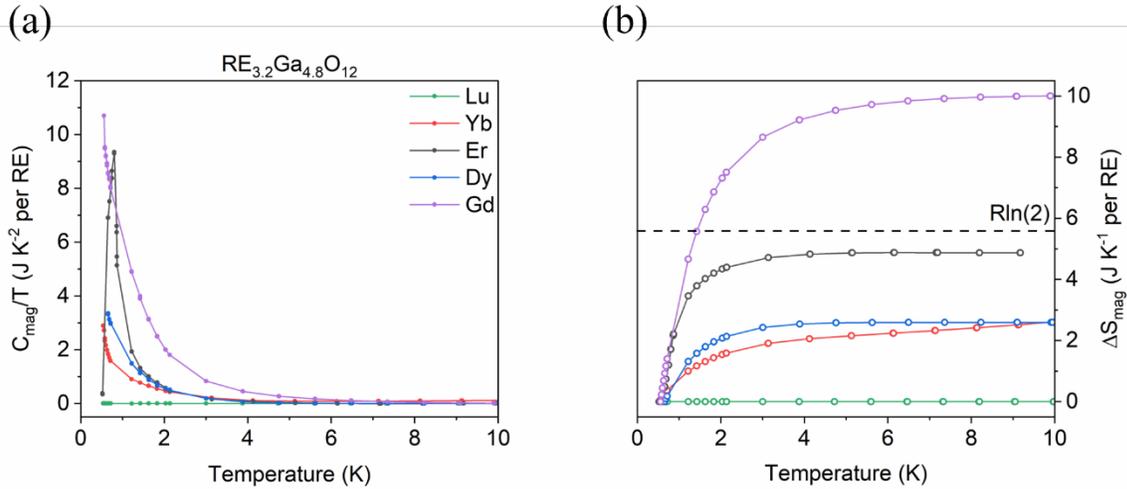

**Figure 6. low temperature characterization of the stuffed garnets.** The specific heat capacity divided by the temperature (a) and the magnetic entropy (b) (obtained by fitting the higher temperature heat capacity to the Debye law and then subtracting the result from the observation) for RE$_{3.2}$Ga$_{4.8}$O$_{12}$ under zero field. R ln 2 is shown as a dashed line in panel b for the Ising model.

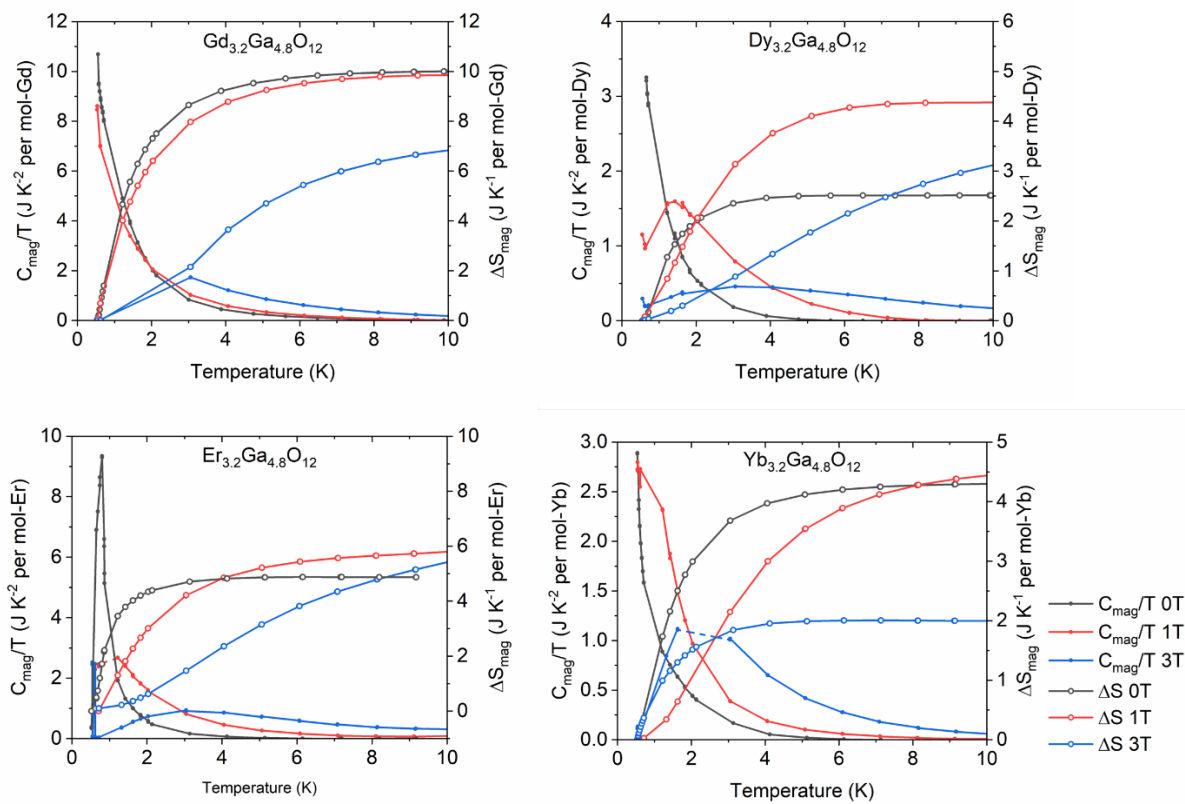

**Figure 7.** Temperature-dependent specific heat capacity data under different applied magnetic fields for the stuffed magnetic garnets, RE$_{3.2}$Ga$_{4.8}$O$_{12}$. Dashed lines are for the guide of the eye.

**Conclusions**

We have presented the synthesis of polycrystalline rare earth excess gallium garnet samples at 1500 C in air. The PXRD patterns showed that the rare earth ionic radius affects the maximum possible stuffing concentration. For a future study on stuffed garnets, a different synthesis method can be tried. A more comprehensive magnetism and spin dynamic study can be done by collecting neutron scattering, muon spectroscopy, or other data sensitive to the rare earth magnetic state. Changes in the magnetic anisotropy and the local magnetic ordering in stuffed garnets would be of interest. Researchers interested in stuffed systems can potentially derive a crystal field model for the stuffed garnets and compare it with the experimental data. This study generally characterizes the stuffed rare earth garnets; researchers interested in the specific study of one or more of them can use the information presented here as a baseline.

**Acknowledgments**

This material is based upon work supported by the U.S. Department of Energy, Office of Science, National Quantum Information Science Research Centers, Co-design Center for Quantum Advantage (C2QA) under contract number DE-SC0012704. The research at Michigan State University was funded by the U.S. Department of Energy, Basic Energy Sciences under contract number DE-SC0023568.

**Associated Contents**

Supporting Information Available

The supporting information contains magnetic susceptibilities under magnetic fields, and the magnetic entropy for the full Yb series. This information is available free of charge at the website: http://pubs.acs.org/

# References

(1) Lau, G. C.; McQueen, T. M.; Huang, Q.; Zandbergen, H. W.; Cava, R. J. Long- and Short-Range Order in Stuffed Titanate Pyrochlores. *J. Solid State Chem.* **2008**, *181* (1), 45–50. https://doi.org/10.1016/j.jssc.2007.10.025.

(2) Dunsiger, S. R.; Gardner, J. S.; Chakhalian, J. A.; Cornelius, A. L.; Jaime, M.; Kiefl, R. F.; Movshovich, R.; MacFarlane, W. A.; Miller, R. I.; Sonier, J. E.; Gaulin, B. D. Low Temperature Spin Dynamics of the Geometrically Frustrated Antiferromagnetic Garnet ${\mathrm{Gd}}_{3}{\mathrm{Ga}}_{5}{O}_{12}$. *Phys. Rev. Lett.* **2000**, *85* (16), 3504–3507. https://doi.org/10.1103/PhysRevLett.85.3504.

(3) Bramwell, S. T.; Gingras, M. J. P. Spin Ice State in Frustrated Magnetic Pyrochlore Materials. *Science* **2001**, *294* (5546), 1495–1501. https://doi.org/10.1126/science.1064761.

(4) Bramwell, S. T.; Harris, M. J. The History of Spin Ice. *J. Phys. Condens. Matter* **2020**, *32* (37), 374010. https://doi.org/10.1088/1361-648X/ab8423.

(5) Dukhovskaya, E. L.; Saksonov, Yu. G.; Titova, A. G. Oxygen Parameters of Certain Compounds of the Garnet Structure. *Izv. Akad. Nauk SSSR Neorganicheskie Mater.* **1973**, *9* (5), 809–813.

(6) Paddison, J. A. M.; Jacobsen, H.; Petrenko, O. A.; Fernández-Díaz, M. T.; Deen, P. P.; Goodwin, A. L. Hidden Order in Spin-Liquid Gd $_3$ Ga $_5$ O $_{12}$. *Science* **2015**, *350* (6257), 179–181. https://doi.org/10.1126/science.aaa5326.

(7) Khatua, J.; Bhattacharya, S.; Ding, Q. P.; Vrtnik, S.; Strydom, A. M.; Butch, N. P.; Luetkens, H.; Kermarrec, E.; Rao, M. S. R.; Zorko, A.; Furukawa, Y.; Khuntia, P. Spin Liquid State in a Rare-Earth Hyperkagome Lattice. *Phys. Rev. B* **2022**, *106* (10), 104404. https://doi.org/10.1103/PhysRevB.106.104404.

(8) Lau, G. C.; Freitas, R. S.; Ueland, B. G.; Dahlberg, M. L.; Huang, Q.; Zandbergen, H. W.; Schiffer, P.; Cava, R. J. Structural Disorder and Properties of the Stuffed Pyrochlore Ho $_2$ Ti O $_5$. *Phys. Rev. B* **2007**, *76* (5), 054430. https://doi.org/10.1103/PhysRevB.76.054430.

(9) Lau, G. C.; Muegge, B. D.; McQueen, T. M.; Duncan, E. L.; Cava, R. J. Stuffed Rare Earth Pyrochlore Solid Solutions. *J. Solid State Chem.* **2006**, *179* (10), 3126–3135. https://doi.org/10.1016/j.jssc.2006.06.007.

(10) Aparnadevi, N.; Saravana Kumar, K.; Manikandan, M.; Santhosh Kumar, B.; Stella Punitha, J.; Venkateswaran, C. Structural Properties, Optical, Electrical and Magnetic Behavior of Bismuth Doped Gd3Fe5O12 Prototype Garnet. *J. Mater. Sci. Mater. Electron.* **2020**, *31* (3), 2081–2088. https://doi.org/10.1007/s10854-019-02729-4.

(11) Borodin, V. A.; Doroshev, V. D.; Tarasenko, T. N.; Savosta, M. M.; Novak, P. Magnetic Structure and Modification of Hyperfine Field in ErxY3-XFe5O12garnets: A Nuclear Magnetic Resonance Study. *J. Phys. Condens. Matter* **1991**, *3* (31), 5881–5892. https://doi.org/10.1088/0953-8984/3/31/011.

(12) Zorenko, Yu.; Gorbenko, V.; Zorenko, T.; Vasylkiv, Ya. Luminescent Properties of the Sc3+ Doped Single Crystalline Films of (Y,Lu,La)3(Al,Ga)5O12 Multi-Component Garnets. *Opt. Mater.* **2014**, *36* (10), 1760–1764. https://doi.org/10.1016/j.optmat.2014.03.028.

(13) Baran, M.; Kissabekova, A.; Krasnikov, A.; Reszka, A.; Vasylechko, L.; Zazubovich, S.; Zhydachevskyy, Y. Luminescence and Defects Creation Processes in Bi3+-Doped Gallium Garnets. *J. Lumin.* **2023**, *253*, 119483. https://doi.org/10.1016/j.jlumin.2022.119483.

(14) Xu, J.; Ueda, J.; Kuroishi, K.; Tanabe, S. Fabrication of Ce3+–Cr3+ Co-Doped Yttrium Aluminium Gallium Garnet Transparent Ceramic Phosphors with Super Long Persistent Luminescence. *Scr. Mater.* **2015**, *102*, 47–50. https://doi.org/10.1016/j.scriptamat.2015.01.029.

(15) Charnaya, E. V.; Shevchenko, E. V.; Khazanov, E. N.; Taranov, A. V.; Ulyashev, A. M. Features of the Low-Temperature Heat Capacity of Er3 –XTmxAl5O12 Garnet Single Crystals. *J. Commun. Technol. Electron.* **2019**, *64* (8), 811–817. https://doi.org/10.1134/S1064226919070064.

(16) Ross, K. A.; Proffen, Th.; Dabkowska, H. A.; Quilliam, J. A.; Yaraskavitch, L. R.; Kycia, J. B.; Gaulin, B. D. Lightly Stuffed Pyrochlore Structure of Single-Crystalline Yb 2 Ti 2 O 7 Grown by the Optical Floating Zone Technique. *Phys. Rev. B* **2012**, *86* (17), 174424. https://doi.org/10.1103/PhysRevB.86.174424.

(17) Zhou, H. D.; Wiebe, C. R.; Jo, Y. J.; Balicas, L.; Qiu, Y.; Copley, J. R. D.; Ehlers, G.; Fouquet, P.; Gardner, J. S. The Origin of Persistent Spin Dynamics and Residual Entropy in the Stuffed Spin Ice Ho2.3Ti1.7O7−δ. *J. Phys. Condens. Matter* **2007**, *19* (34), 342201. https://doi.org/10.1088/0953-8984/19/34/342201.

(18) *Highly Nonstoichiometric YAG Ceramics with Modified Luminescence Properties - Cao - 2023 - Advanced Functional Materials - Wiley Online Library*. https://onlinelibrary.wiley.com/doi/full/10.1002/adfm.202213418 (accessed 2023-08-31).

(19) Yang, C.; Wang, H.; Jin, L.; Xu, X.; Ni, D.; Thompson, J. D.; Xie, W.; Cava, R. J. Erbium-Excess Gallium Garnets. *Inorg. Chem.* **2023**. https://doi.org/10.1021/acs.inorgchem.3c01132.

(20) Yang, C.; Wang, H.; Jin, L.; Xu, X.; Ni, D.; Thompson, J. D.; Xie, W.; Cava, R. J. Erbium-Excess Gallium Garnets. arXiv June 20, 2023. https://doi.org/10.48550/arXiv.2306.11854.

(21) Sana, B.; Barik, M.; Jena, U.; Baenitz, M.; Sichelschmidt, J.; Sethupathi, K.; Khuntia, P. Magnetic Properties of a Spin-Orbit Entangled Jeff=1/2 Three-Dimensional Frustrated Rare-Earth Hyperkagome. arXiv April 14, 2023. http://arxiv.org/abs/2304.07350 (accessed 2023-06-16).

(22) Mugiraneza, S.; Hallas, A. M. Tutorial: A Beginner's Guide to Interpreting Magnetic Susceptibility Data with the Curie-Weiss Law. *Commun. Phys.* **2022**, *5* (1), 1–12. https://doi.org/10.1038/s42005-022-00853-y.

(23) Lhotel, E.; Mangin-Thro, L.; Ressouche, E.; Steffens, P.; Bichaud, E.; Knebel, G.; Brison, J.-P.; Marin, C.; Raymond, S.; Zhitomirsky, M. E. Spin Dynamics of the Quantum Dipolar Magnet Yb 3 Ga 5 O 12 in an External Field. *Phys. Rev. B* **2021**, *104* (2), 024427. https://doi.org/10.1103/PhysRevB.104.024427.

(24) Cai, Y.; Wilson, M. N.; Beare, J.; Lygouras, C.; Thomas, G.; Yahne, D. R.; Ross, K.; Taddei, K. M.; Sala, G.; Dabkowska, H. A.; Aczel, A. A.; Luke, G. M. Crystal Fields and Magnetic Structure of the Ising Antiferromagnet Er 3 Ga 5 O 12. *Phys. Rev. B* **2019**, *100* (18), 184415. https://doi.org/10.1103/PhysRevB.100.184415.

(25) Marshall, I. M.; Blundell, S. J.; Pratt, F. L.; Husmann, A.; Steer, C. A.; Coldea, A. I.; Hayes, W.; Ward, R. C. C. A Muon-Spin Relaxation (MSR) Study of the Geometrically Frustrated Magnets Gd3Ga5O12 and ZnCr2O4. *J. Phys. Condens. Matter* **2002**, *14* (6), L157. https://doi.org/10.1088/0953-8984/14/6/104.

(26) Schiffer, P.; Ramirez, A. P.; Huse, D. A.; Valentino, A. J. Investigation of the Field Induced Antiferromagnetic Phase Transition in the Frustrated Magnet: Gadolinium Gallium

Garnet. *Phys. Rev. Lett.* **1994**, *73* (18), 2500–2503. https://doi.org/10.1103/PhysRevLett.73.2500.
(27) Daudin, B.; Lagnier, R.; Salce, B. Thermodynamic Properties of the Gadolinium Gallium Garnet, Gd3Ga5O12, between 0.05 and 25 K. *J. Magn. Magn. Mater.* **1982**, *27* (3), 315–322. https://doi.org/10.1016/0304-8853(82)90092-0.
(28) Summers, L. T. *Advances in Cryogenic Engineering Materials*; Springer, 2013.
(29) Gómez-Coca, S.; Urtizberea, A.; Cremades, E.; Alonso, P. J.; Camón, A.; Ruiz, E.; Luis, F. Origin of Slow Magnetic Relaxation in Kramers Ions with Non-Uniaxial Anisotropy. *Nat. Commun.* **2014**, *5* (1), 4300. https://doi.org/10.1038/ncomms5300.
(30) Kimura, H.; Maeda, H.; Sato, M. Single Crystals Growth and Magneto-Thermal Properties of Dy3Ga5O12 Garnet. *J. Mater. Sci.* **1988**, *23* (3), 809–813. https://doi.org/10.1007/BF01153971.